# Active Huygens' metasurface based on in-situ grown conductive polymer

*Wenzheng Lu[1,*], Leonardo de S. Menezes[1,2,3], Andreas Tittl[1], Haoran Ren[4] and Stefan A. Maier[4,5,*]*

[1]*Chair in Hybrid Nanosystems, Nano-Institute Munich, Faculty of Physics, Ludwig-Maximilians-Universität München, Munich, 80539, Germany.*

[2]*Departamento de Física, Universidade Federal de Pernambuco, 50670-901 Recife-PE, Brazil.*

[3]*Center for Nanoscience, Faculty of Physics, Ludwig-Maximilians-Universität München, Munich, 80539, Germany.*

[4]*School of Physics and Astronomy, Monash University, Clayton, Victoria 3800, Australia.*

[5]*Department of Physics, Imperial College London, London SW72AZ, UK.*

[*]*e-mail: Wenzheng.Lu@physik.uni-muenchen.de; Stefan.Maier@monash.edu*

**Abstract**

Active metasurfaces provide unique advantages for on-demand light manipulation at a subwavelength scale for emerging applications of 3D displays, augmented/virtual reality (AR/VR) glasses, holographic projectors and light detection and ranging (LiDAR). These applications put stringent requirements on switching speed, cycling duration, controllability over intermediate states, modulation contrast, optical efficiency and operation voltages. However, previous demonstrations focus only on particular subsets of these key performance requirements for device implementation, while the other performance metrics have remained too low for any practical use. Here, we demonstrate an active Huygens' metasurface based on in-situ grown conductive polymer with holistic switching performance, including switching speed of 60 frames per second (fps), switching duration of more than 2000 switching cycles without noticeable degradation, hysteresis-free controllability over intermediate states, modulation contrast of over 1400%, optical efficiency of 28% and operation voltage range within 1 V. Our active metasurface design meets all foundational requirements for display applications and can be readily incorporated into other metasurface concepts to deliver high-reliability electrical control over its optical response, paving the way for compact and robust electro-optic metadevices.

**Introduction**

Active metasurfaces, often alternatively termed as tunable or reconfigurable metasurfaces, are rapidly emerging as a major frontier in photonic research and have launched tremendous breakthroughs in modern optics[1–3]. Compared with their passive counterparts, active metasurfaces consist of ultrathin planar arrays of subwavelength active nanoantennas, whose optical responses can be dynamically modulated on-demand. Over the past decade, active metasurfaces have made a significant impact on the new development of beam deflectors[4–8], spatial light modulators[9–12], varifocal metalenses[13–15], dynamic holograms[16–19] and many others. The active tuning schemes mainly rely on varying optical properties of the nanoantennas or their surrounding materials through chemical reactions[19–21], mechanical displacements[22–24], electrical switching[16,17,25–27], thermal modulation[28–31] and all-optical switching[32–34]. Among these modulation schemes, electrical switching is of particular interest because it promises compact integration of meta-optics with miniaturized on-chip electro-optic systems, which can be readily incorporated into electronic smart devices for practical applications[1,2], such as 3D displays, augmented/virtual reality (AR/VR) glasses, dynamic holograms, beam steering and light detection and ranging (LiDAR).

To this end, a number of active metasurfaces have been implemented by using different electro-active materials, such as chalcogenide phase-change materials (PCMs)[26–27], III-V semiconducting materials[5,35,36], ionic conducting materials[37–38], metallic polymers[6,14]. Previous active metasurfaces exhibit only distinct subsets of key performance metrics[1], including switching speed, cycling duration, controllability over intermediate states, modulation contrast, optical efficiency and operation voltages, as shown in Figure 1a. However, these performance attributes are hardly met simultaneously, making them almost impossible for any practical use. For instance, even though PCM metasurfaces can provide a switching speed of up to 2 MHz[27], they require high operation voltage and large cell manipulation[26]. III-V semiconductors offer excellent switching duration and controllability for intermediate states, but suffer from low modulation contrasts due to the small tuning range of the material intrinsic refractive index[5]. Recently, conductive polymers have shown many desired properties for electrically active metasurfaces, such as large variation of refractive index, fast switching speed, superior cycling stability and low operation voltages[39–42]. However, previous active metasurfaces purely based on conductive polymers are disadvantaged by low diffractive efficiency, due to the geometric phase design and the weak resonant nature of the nanostructured polymer[6,14]. Moreover, the fabrication of the conductive polymer relying on spin-coating prevents long-term switching durability since the contact issue between the polymer and the substrate[6,14].

In this work, we introduce an electrically active Huygens' metasurface based on the in-situ grown conductive polymer, polyaniline (PANI), and experimentally demonstrate its holistic active switching performance. We combine the superior electro-optical response of PANI with the accurate and continuous phase modulation of the Huygens' nanoantennas, and achieve an unprecedented switching performance covering all key requirements for display applications. Our active Huygens' metasurface exhibits a high modulation contrast of 1400% and a diffraction efficiency up to 28%, which is 25 times higher than the previous polymer-based metasurface[6,14] and comparable to liquid crystal-based metadevices[9]. The manufacture of the active PANI layer is realized by in-situ polymer growth on the pre-fabricated metasurface. The solid contact between the nanoantennas and the grown PANI facilitates mechanical and electrical durability, enabling a superior cycling duration of over 2000 switching cycles without noticeable degradations. The intrinsic dynamic properties of PANI endow the active metasurface with fast switching speed of 60 fps and virtually hysteresis-free controllability over continuous intermediate states within a low operation voltage range from -0.2 V to +0.6 V. Unlike previous active metasurfaces performing well only in subsets of performance metrics (Fig. 1a), our electrically active metasurface with holistic switching performance holds great promise towards practical display applications.

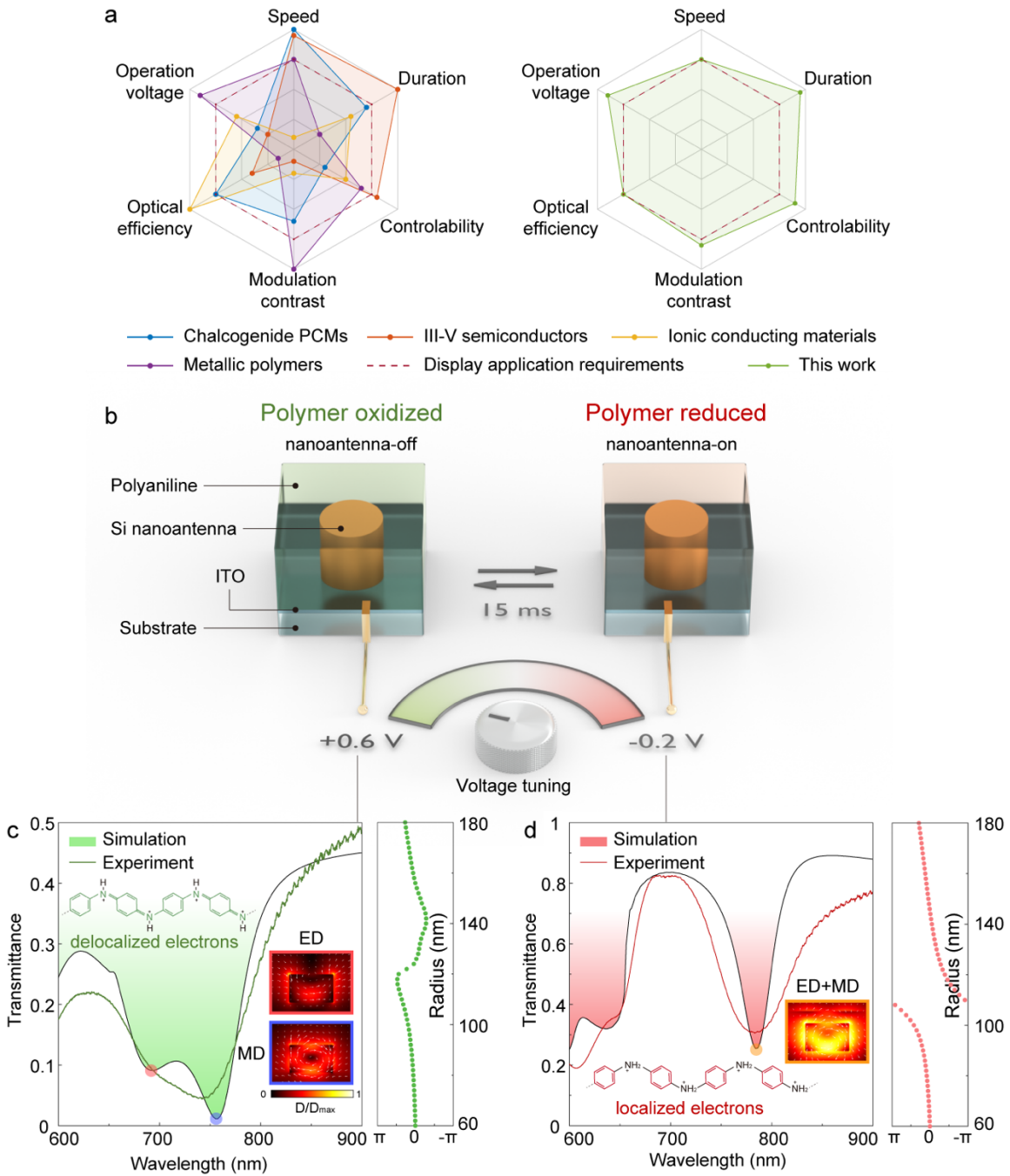

**Fig. 1 | Comparison of active metasurface performance and the concept of electrically active Huygens' metasurface composed of PANI-modulated dielectric nanoantennas. a**, Comparison of different electrically active metasurfaces in performances. The performance is evaluated relatively to the display application requirements defined in the reference[1]. The evaluation of each

electro-active materials is based on representative references: chalcogenide PCMs[26,27], III-V semiconductors[5,36], ionic conducting materials[37,38], metallic polymers[6,14]. **b**, Schematics illustrating the electrical switching of an individual active Huygens' nanoantenna based on conductive polymer. The polymer is switched to oxidized state at +0.6 V (left), and to reduced state at -0.2 V (right). The switching time between the states is 15 ms, corresponding to a refresh rate of over 60 fps. **c**, Simulated and measured transmission spectra (left panel) of a metasurface made of homogeneous nanoantennas (height $H$ = 140 nm, radius $R$ = 110 nm, periodicity $P_x = P_y$ = 450 nm) and the calculated phase retardation (right panel) of the nanoantennas with different radii in the oxidized state. **d**, Simulated and measured transmission spectra (left panel) and the calculated phase retardation (right panel) in the reduced state. The inset chemical structures show the conversion of the polymer between the oxidized and reduced states. The inset images show the simulated field distributions of the electric dipole (ED, red) and the magnetic dipole (MD, blue) in the oxidized state, as well as the spectrally overlapping electric and magnetic dipoles (ED + MD, yellow) in the reduced state, at their corresponding resonant wavelengths of 694 nm, 765 nm and 785 nm, respectively.

**Results**

**Concept of electrically active Huygens' nanoantennas**

Our active Huygens' metasurface is composed of dielectric silicon nanodisks surrounded by a layer of PANI, resting on an indium-tin-oxide (ITO) substrate that acts as an electrical contact (Fig. 1b). The cylindrical shape of the nanoantenna is chosen to provide polarization independence and the period of the nanoantennas is fixed to 450 nm in both $x$ and $y$ direction so that the unit cell is subdiffractive. The state of the polymer is collectively controlled by voltage tuning via electrochemistry, where an applied voltage range between +0.6 V and -0.2 V (vs. a reference electrode) can continuously tune the polymer into states in between the oxidized and reduced states. In detail, the redox reaction triggered by the applied voltage induces an alternating delocalization of the π-electrons in the polymer chain of PANI, resulting in a large variation of the refractive index with a maximum $\varDelta n$ = 0.6 at 780 nm (Supplementary Fig. 1).

The core concept of our design is the spectral tuning of the electric dipole (ED) and magnetic dipole (MD) of the nanoantenna by electrical modulation of the surrounding polymer. Fig. 1c and d depict the spectral response of nanoantennas at different applied voltages. In the oxidized state, the simulated and measured transmission spectra show that the ED and the MD are spectrally separated, resonating at 694 nm and 765 nm, respectively. When switched to the reduced state, the two dipole resonances redshift to 785 nm and overlap with each other. As suggested by the principle of dielectric Huygens' metasurfaces[43], the spectrally overlapping resonances of ED and MD signifies the fulfillment of Huygens' condition, which produces a full-range phase shift from 0 to 2π combined with high transmission. Importantly, the phase modulation functionality of the nanoantenna is switched on at an applied voltage of -0.2 V, exhibiting strong phase shift (Fig. 1d), while at +0.6 V, the nanoantenna exhibits a phase retardation well below 2π-phase coverage and can be considered switched off (Fig. 1c). The variation of the phase shift in the nanoantennas is determined by the interference of the ED and MD resonances, where the spectral responses of the two resonances are determined by the nanoantenna geometrics and the refractive-index contrast between the nanoantenna and the surrounding environment[43]. Therefore, by harnessing the environmental tuning scheme and carefully choosing nanoantenna geometries for optimum phase engineering, switchable metasurfaces with high efficiency and greatly improved switching contrast can be realized.

**Metasurface design for electrical beam steering**

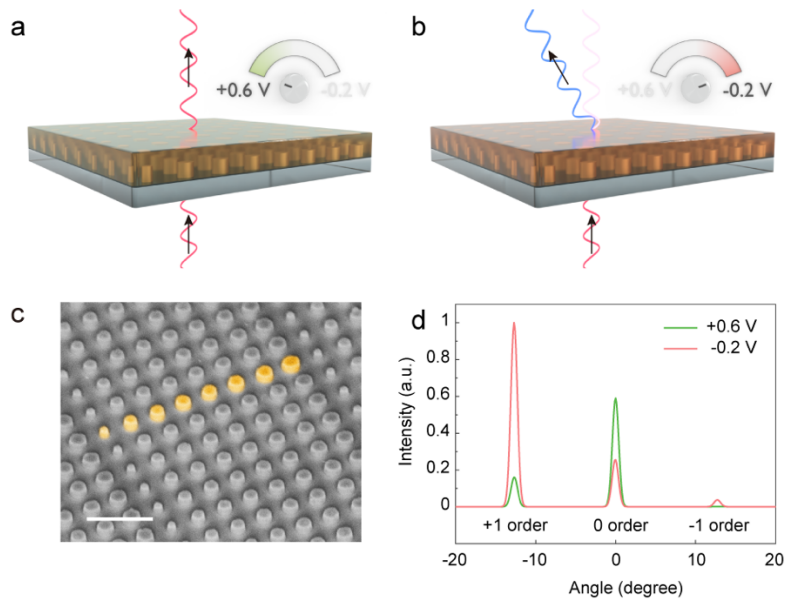

**Fig. 2 | Metasurface design for electrical beam steering. a,b**, Schematic illustrating the concept of electrical beam steering using a polymer-integrated Huygens' metasurface at the applied voltage of +0.6 V and -0.2 V, respectively. **c**, Tilted SEM image of the bare dielectric metasurface with eight-element gradient nanoantennas design (highlighted structures). Scale bar: 1 μm. **d**, Simulated far-field transmitted intensities of +1, 0, -1 diffraction orders for the electrical beam steering metasurface.

As a proof of concept, we implemented an active metasurface for optical beam steering, where the incident light can be deflected into a fixed angle on-demand by electrical control (Fig. 2a and b). The metasurface design is based on an eight-element gradient phase geometry using the calculated nanoantenna phase profiles for an incident wavelength of 785 nm (Supplementary Fig. 2). The chosen structures have an average transmittance of 0.6. The metasurfaces were fabricated via electron-beam lithography (EBL) patterning and reactive-ion etching (RIE), followed by electrochemical polymerization for the growth of the polymer. Figure 2c shows a scanning electron microscopy (SEM) image of the fabricated bare dielectric metasurface on a transparent conductive ITO/glass substrate. The metasurface is designed such that the steered beam couples to the +1 diffraction order at the polymer reduced state, and to the 0 order at the polymer oxidized state. The design principle is validated by simulated electric field distributions (Supplementary Fig. 3), where the transmitted spatial phase profile supports diffraction at +1 order at an operation voltage of -0.2 V and strongly suppresses the diffraction at +0.6 V. Detailed far-field analysis of the metasurfaces reveals the switchable beam steering performance with high switching contrast at a diffraction angle of 12.7°, meanwhile the intensity of the 0 order can be readily modulated (Fig. 2d).

**In-situ polymer growth and electrical beam steering performance.**

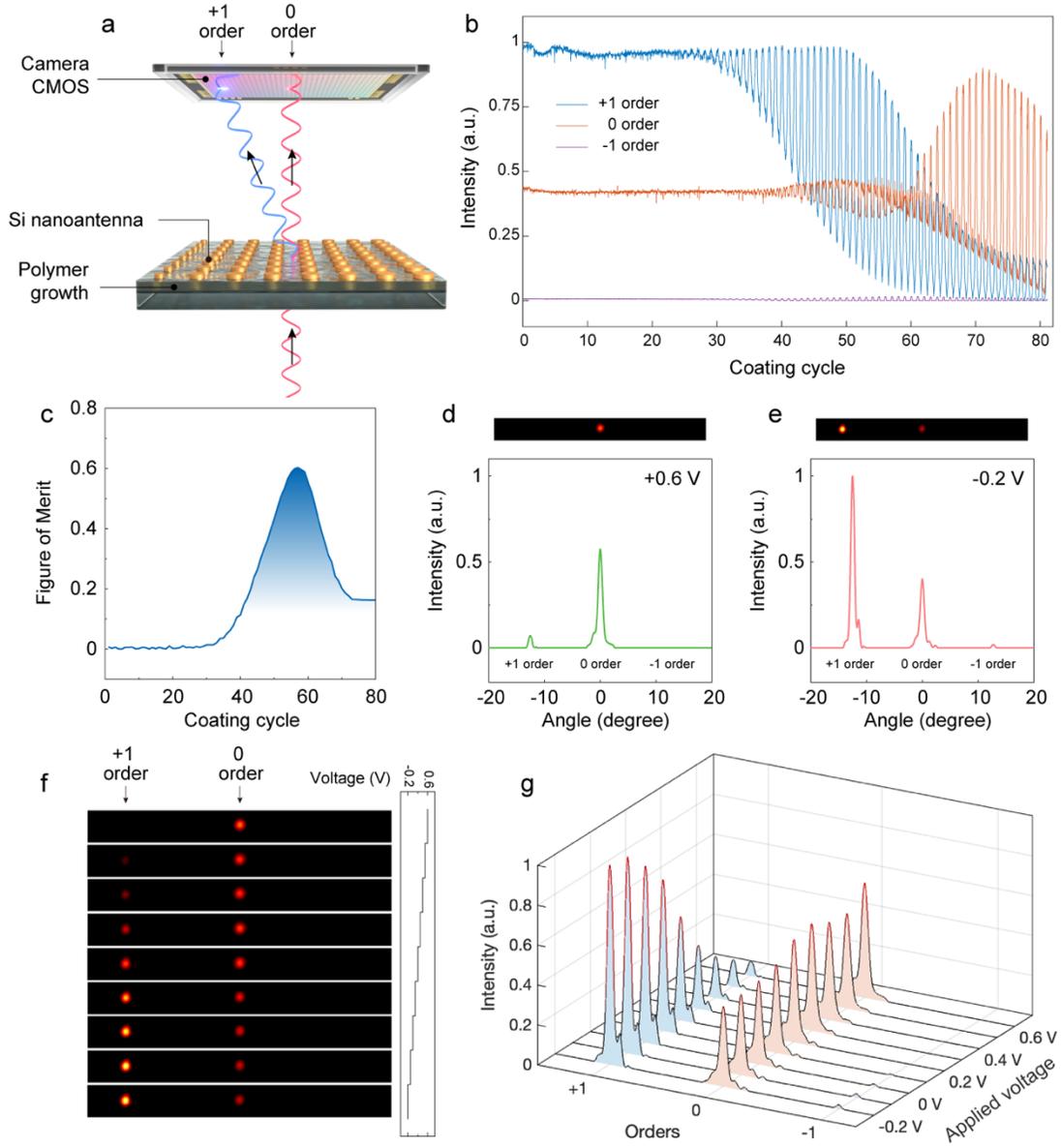

**Fig. 3 | In-situ measurement and optimization of beam steering performance. a**, Schematic of the in-situ optical measurement setup for the active beam steering metasurface. A high-speed monochromatic camera was used to record the diffraction pattern during the polymer growth process. **b**, Normalized intensities of +1, 0 and -1 diffraction orders during the polymer growth. **c**, Figure of merit (FOM) of beam steering between 0 and +1 diffraction orders at different coating cycles obtained from the measured intensities in **b**. **d**, Measured camera image (top) and diffraction

intensity profile (bottom) at an applied voltage of +0.6 V. **e**, Measured camera image (top) and diffraction intensity profile (bottom) at an applied voltage of -0.2 V. **f**, Camera images of the diffraction orders at different input voltages. **g**, Diffraction intensity profiles at different applied voltage extracted from **f**. The data in **d-f** were measured on the beam steering metasurface at an optimized polymer coating cycle number of 56.

To experimentally validate our electrical beam steering metasurfaces, electrochemical growth of PANI was implemented on the fabricated metasurface in-situ in the optical measurement setup, where a high-speed monochromatic camera was used to monitor the diffraction pattern during the polymer growth (Fig. 3a). A three-electrode system integrated with the optical measurement setup was employed for PANI growth and electrical switching in an aqueous electrolyte using a custom-built electrochemical cell (Supplementary Fig. 4, see also Methods for further information). Since PANI is switched between -0.2 V and +0.6 V and can be electrochemically polymerized under an applied voltage of +0.8 V on an ITO electrode[44–45], a cyclic voltammetry with a voltage range from -0.2 V to +0.8 V was chosen to carry out PANI growth and electrical switching simultaneously (Supplementary Fig. 5, see also Supplementary Note 1), thus allowing for in-situ optimization of the electrical beam steering performance.

The diffraction images were monitored throughout the entire polymer growth process. Figure 3b shows the intensities of +1, 0 and -1 diffraction orders as function of the PANI coating cycle. In the first 25 coating cycles, the variation of the diffractive intensities is very small due to the insufficient amount of the grown PANI. By increasing the coating cycle number and thus the PANI thickness, the effect of the grown PANI layer becomes large enough to enable switching of the diffraction intensities. As predicted by the design, the diffraction intensity of the +1 order reaches its minima and maxima at the oxidized state and the reduced state, respectively, while the 0 order shows an opposite response. The intensity contrasts of the +1 and 0 diffraction orders at the two states are dictated by the optical phase variation and the intrinsic optical absorption of PANI. Further increasing the PANI thickness allows for the optical absorption of PANI to become more influential, as evidenced by the decreased diffractive intensity and switching contrast in the +1 order. The overcoated PANI also leads to a phase mismatch in the nanoantennas, where the switching of the +1 and 0 diffraction orders becomes synchronous and solely dependent on variation of optical absorption of PANI.

Quantitative evaluation of the electrical beam steering performance is accomplished by a figure-of-merit (FOM) generally used for beam deflectors, which is based on the switching contrasts between the +1 and 0 diffraction orders at two states[28]. The FOM can be represented as:

$$FOM = \frac{1}{2}(C^{ox} - C^{red})$$
$$= \frac{1}{2}\left(\frac{I_0^{ox} - I_1^{ox}}{I_0^{ox} + I_1^{ox}} - \frac{I_0^{red} - I_1^{red}}{I_0^{red} + I_1^{red}}\right) \quad (1)$$

where $C$ is the switching contrast between two diffraction orders, $I_0$ and $I_1$ represent the intensities of the 0 and +1 diffraction orders, respectively, and the superscripts of *ox* and *red* denote the oxidized and reduced states, respectively. The value of FOM, which falls between 0 and 1, describes the capability of a beam deflector to route energy into certain angles, where FOM = 1 represents the ideal case. The calculated FOM for different coating cycles indicates that the beam steering performance is optimized when the FOM reaches 0.6 with a polymer coating cycle number of 56 (Fig. 3c). In comparison, the obtained FOM from a chalcogenide PCM-based beam steering metasurface[26] is 0.42, and 0.38 from a liquid crystal-based one[9]. Figure 3d and e present the diffraction images at the optimized thickness and the corresponding intensity profiles, at +0.6 V and -0.2 V, respectively, and the results are in excellent agreement with the simulation. The modulation contrast, defined as the intensity ratio of the +1 order between the two states, is over 1400% and can be further increased by increasing the PANI thickness. The thickness of the PANI layer at the optimized condition is measured to be 200 nm (Supplementary Fig. 6), which is only 1/7 as thin as the thickness of the embedding active layer in liquid crystal-based active metasurfaces[9–17]. So far, the diffraction efficiency of the +1 order at the reduced state is 28.8%, where the efficiency loss induced by the optical loss of the grown PANI is 3.7%. In comparison with the beam steering metasurfaces purely based on conductive polymers[6,14], the diffraction efficiency of our design is 25 times higher, being even comparable to the liquid crystal-based metadevices[9]. However, the optical loss of the oxidized PANI is noticeable and accounts for up to 70% of the transmitted energy of the reduced state (Supplementary Fig. 7). The diffraction efficiency can be further increased by decreasing the polymer thickness together with optimizations of the nanoantenna geometry and materials, allowing to compensate the optical loss of the polymer.

In addition, intermediate PANI states, which possess gradually changing refractive indices due to the partially delocalized π-electrons, can be accessed via voltage tuning, thus allowing for a continuous tuning of the phase of the Huygens' nanoantennas. By operating the voltage between the oxidized and reduced states, we can modify on-demand the diffraction intensity and the intensity ratio between the +1 and 0 diffraction orders, as demonstrated in the diffraction images (Fig. 3f) and the intensity profiles (Fig. 3g) at different applied voltages. Notably, the electrical switching demonstrates high-quality reversibility. With a cycling voltage between -0.2 V and +0.6 V, the electrical switching is fully reversible with a remarkable stability on our active metasurface, as shown in the Supplementary Video 1 which includes the diffraction images, diffraction intensities and the cyclic voltammograms for a total of 9 switching cycles in real-time. We further analyzed the hysteresis behavior of the electrical switching (Supplementary Fig. 8). Our electrically active metasurfaces can be operated in a nearly hysteresis-free manner. This allows to switch to the intermediate states without a memory effect that could possibly deteriorate the accuracy of optical response by electrical modulation, which is crucial for realizing active metasurfaces for precise on-demand control of phase and amplitude.

**Electrical switching speed and durability**

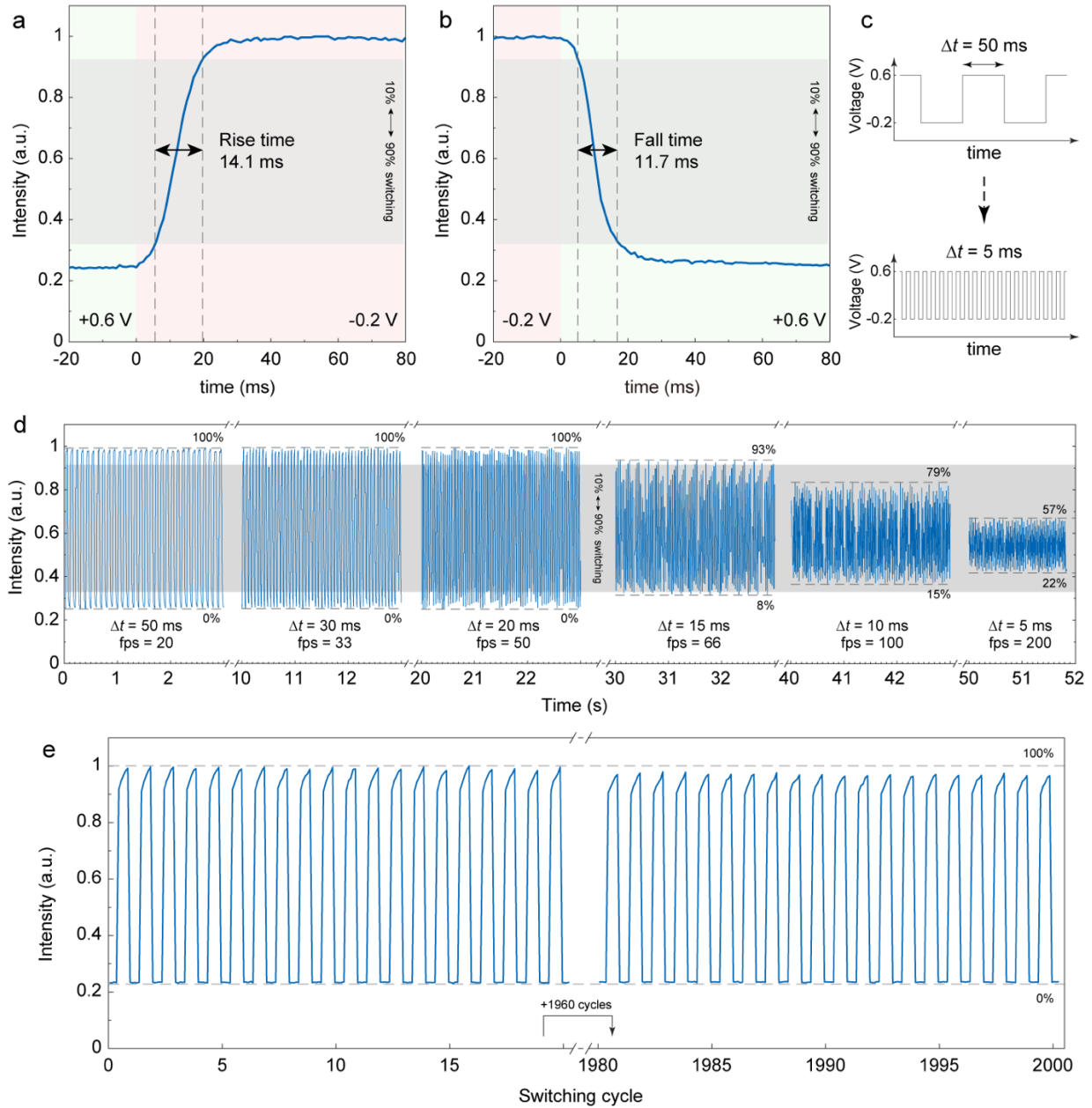

**Fig. 4 | Electrical switching speed and durability. a,b**, Temporal responses for the switch-on and switch-off processes, respectively. The rise time ($\tau_{rise}$) of the switch-on process is 14.1 ms and the fall time ($\tau_{fall}$) of the switch-off process is 11.7 ms. The rise time and the fall time are defined as the time required to switch between 10% and 90% of the optical response, depicted as the gray area, under an abrupt alternation of the applied voltage (green and red areas). **c**, Input voltage schemes with different voltage duration times ($\Delta t$) in a range from 50 ms to 5 ms. **d**, Temporal

optical response at different input voltage schemes. Each scheme lasts 10 s. The percentage numbers indicate the maximum accessible range of the optical response. The recorded slightly inconsistent maxima and minima are due to frame drops of the camera. **e**, Switching durability of the electrically active metasurface under $\Delta t$ = 3 s. The electrical switching is highly reversible and consistent without noticeable degradation over 2000 switching cycles. The optical response in this figure corresponds to the +1 order at an optimized coating cycle of PANI.

To evaluate the switching speed of our active metasurfaces, the temporal optical response is measured under an abrupt alternation of the input voltages between the on (-0.2 V) and off (+0.6 V) states, as presented in Figure 4a and b. The rise time ($\tau_{rise}$) of the switch-on process and the fall time ($\tau_{fall}$) of the switch-off process, defined as the time required for the modulated intensity to rise or fall between the 10% and 90% switching window, are 14.1 ms and 11.7 ms, respectively. We further demonstrated the switching speed can be utilized for electro-optic metadevices under a high-speed switching configuration. The metasurface is operated at alternating input voltages between the on and off states with a voltage duration time ($\Delta t$) varied from 50 ms to 5 ms (Fig. 4c). Such intense switching conditions could possibly lead to mechanical or electrical failures due to the volume expansion, physical exfoliation or irreversible reactions of polymer films[46]. However, remarkably, when $\Delta t$ approaches the switching time $\tau_{rise}$ and $\tau_{fall}$, the accessible range of the optical response approaches the 10% to 90% switching window at $\Delta t$ = 15 ms, equivalent to 66 fps, as shown in Figure 4d. Lowering $\Delta t$ below 15 ms decreases the accessible optical range, but does not significantly influence the reversibility. The key characteristic supporting the superior cycling duration is that, when electron injection or extraction takes place by voltage application, the highly conductive PANI is able to redistribute delocalized π-electrons along the polymer chain without structural degradation[47,48]. This ensures the stability when deploying intense high-speed electrical modulation. Moreover, the solid contact between the nanoantennas and the PANI film, which is directly grown on the metasurface, provides additional mechanical and electrical stabilities. The switching durability is further experimentally secured by a long-time low-speed operation (at $\Delta t$ = 3 s), where the electrical switching exhibits no substantial degradations over 2000 switching cycles (Fig. 4e).

**Conclusion**

In this study, we have introduced and experimentally realized an active Huygens' metasurface based on the in-situ grown PANI with holistic switching performance for display applications. We have achieved electrical switching on the beam steering metasurface with large modulation contrast of over 1400%, hysteresis-free controllability, high diffraction efficiency of 28%, fast switching speed of over 60 fps and superior switching duration over 2000 switching cycles, under an operation voltage range only from -0.2 V to +0.6 V. We have demonstrated that the active Huygens' nanoantenna design not only greatly boosts the optical efficiency, but also offers the possibility to continuously tune the phase of the nanoantenna through intermediate states. In addition, we have shown that PANI is an excellent candidate of electro-active materials for active metasurfaces, owing to its large tuning range of refractive index, fast switching speed, outstanding electro-optical properties and superior physical and chemical stability. The in-situ growth of conductive polymer on metasurface is straightforward and more reliable, which can be processed in a low-cost, large-area and scalable manner. Overall, the concept of our electrically active Huygens' nanoantenna design can be broadly extended to a variety of active metasurfaces that are capable of programmable and dynamic wavefront manipulation. From an industrial standpoint, the holistic switching performance, together with the scalable, reliable, cost-effective polymer manufacturability, provide promising opportunities to achieve switchable glasses-free 3D displays, high-resolution AR/VR glasses, LiDAR, and large field-of-view holographic displays, forging the path of active metasurfaces towards commercial success.

**References**

1. Gu, T., Kim, H. J., Rivero-Baleine, C. & Hu, J. Reconfigurable metasurfaces towards commercial success. *Nat. Photon.* **17**, 48–58 (2023).

2. Neshev, D. N. & Miroshnichenko, A. E. Enabling smart vision with metasurfaces. *Nat. Photon.* **17**, 26–35 (2023).

3. Shaltout, A. M., Shalaev, V. M. & Brongersma, M. L. Spatiotemporal light control with active metasurfaces. *Science* **364**, eaat3100 (2019).

4. Yu, J. et al. Electrically tunable nonlinear polaritonic metasurface. *Nat. Photon.* **16**, 72–78 (2022).

5. Wu, P. C. et al. Dynamic beam steering with all-dielectric electro-optic III-V multiple-quantum-well metasurfaces. *Nat. Commun.* **10**, 3654 (2019).

6. Karst, J. et al. Electrically switchable metallic polymer nanoantennas. *Science* **374**, 612–616 (2021).

7. Jong, D. de et al. Electrically switchable metallic polymer metasurface device with gel polymer electrolyte. *Nanophotonics* **12**, 1397–1404 (2023).

8. Abdollahramezani, S. et al. Electrically driven reprogrammable phase-change metasurface reaching 80% efficiency. *Nat. Commun.* **13**, 1696 (2022).

9. Li, S.-Q. et al. Phase-only transmissive spatial light modulator based on tunable dielectric metasurface. *Science* **364**, 1087–1090 (2019).

10. Park, J. et al. All-solid-state spatial light modulator with independent phase and amplitude control for three-dimensional LiDAR applications. *Nat. Nanotechnol.* **16**, 69–76 (2021).

11. Kwon, H., Zheng, T. & Faraon, A. Nano-electromechanical spatial light modulator enabled by asymmetric resonant dielectric metasurfaces. *Nat. Commun.* **13**, 5811 (2022).

12. Shirmanesh, G. K., Sokhoyan, R., Wu, P. C & Atwater, H. A. Electro-optically tunable multifunctional metasurfaces. *ACS Nano* **14**, 6912–6920 (2020).

13. Shalaginov, M. Y. et al. Reconfigurable all-dielectric metalens with diffraction-limited performance. *Nat. Commun.* **12**, 1225 (2021).

14. Karst, J. et al. Electro-active metaobjective from metalenses-on-demand. *Nat. Commun.* **13**, 7183 (2022).

15. Afridi, A. et al. Electrically driven varifocal silicon metalens. *ACS Photonics* **5**, 4497–4503 (2018).

16. Kaissner, R. et al. Electrochemically controlled metasurfaces with high-contrast switching at visible frequencies. *Sci. Adv.* **7**, eabd9450 (2021).

17. Li, J., Yu, P., Zhang, S. & Liu, N. Electrically-controlled digital metasurface device for light projection displays. *Nat. Commun.* **11**, 3574 (2020).

18. Kim, I. et al. Pixelated bifunctional metasurface-driven dynamic vectorial holographic color prints for photonic security platform. *Nat. Commun.* **12**, 3614 (2021).

19. Li, J. et al. Addressable metasurfaces for dynamic holography and optical information encryption. *Sci. Adv.* **4**, eaar6768 (2018).

20. Duan, X., Kamin, S. & Liu, N. Dynamic plasmonic colour display. *Nat. Commun.* **8**, 14606 (2017).

21. Li, J., Chen, Y., Hu, Y., Duan, H. & Liu, N. Magnesium-based metasurfaces for dual-function switching between dynamic holography and dynamic color display. *ACS Nano* **14**, 7892–7898 (2020).

22. Ee, H.-S. & Agarwal, R. Tunable metasurface and flat optical zoom lens on a stretchable substrate. *Nano Lett.* **16**, 2818–2823 (2016).

23. Holsteen, A. L., Cihan, A. F. & Brongersma, M. L. Temporal color mixing and dynamic beam shaping with silicon metasurfaces. *Science* **365**, 257–260 (2019).

24. Li, Q. et al. Metasurface optofluidics for dynamic control of light fields. *Nat. Nanotechnol.* **17**, 1097–1103 (2022).

25. Benea-Chelmus, I.-C. et al. Electro-optic spatial light modulator from an engineered organic layer. *Nat. Commun.* **12**, 5928 (2021).

26. Zhang, Y. et al. Electrically reconfigurable non-volatile metasurface using low-loss optical phase-change material. *Nat. Nanotechnol.* **16**, 661–666 (2021).

27. Wang, Y. et al. Electrical tuning of phase-change antennas and metasurfaces. *Nat. Nanotechnol.* **16**, 667–672 (2021).

28. Komar, A. et al. Dynamic beam switching by liquid crystal tunable dielectric metasurfaces. *ACS Photonics* **5**, 1742–1748 (2018).

29. Liu, X. et al. Thermally dependent dynamic meta-holography using a vanadium dioxide integrated metasurface. *Adv. Opt. Mater.* **7**, 1900175 (2019).

30. Duan, X. et al. Reconfigurable multistate optical systems enabled by $VO_2$ phase transitions. *ACS Photonics* **7**, 2958–2965 (2020).

31. Kepič, P. et al. Optically tunable Mie resonance $VO_2$ nanoantennas for metasurface in the visible. *ACS Photonics* **8**, 1048–1057 (2021).

32. Shcherbakov, M. R. et al. Ultrafast all-optical tuning of direct-gap semiconductor metasurfaces. *Nat. Commun.* **8**, 17 (2017).

33. Leitis, A. et al. All-dielectric programmable Huygens' Metasurfaces. *Adv. Funct. Mater.* **30**, 1910259 (2020).

34. Wang, Q. et al. Optically reconfigurable metasurfaces and photonic devices based on phase change materials. *Nat. Photon.* **10**, 60–65 (2016).

35. Iyer, P. P., Pendharkar, M., Palmstrøm, C. J. & Schuller, J. A. III-V heterojunction platform for electrically reconfigurable dielectric metasurfaces. *ACS Photonics* **6**, 1345–1350 (2019).

36. Lee, J. et al. Ultrafast electrically tunable polaritonic metasurfaces. *Adv. Opt. Mater.* **2**, 1057–1063 (2014).

37. Eaves-Rathert, J. et al. Dynamic color tuning with electrochemically actuated $TiO_2$ metasurfaces. *Nano Lett.* **22**, 1626–1632 (2022).

38. Li, Y. Groep, J. van de, Talin, A. A. & Brongersma, M. L. Dynamic tuning of gap plasmon resonances using a solid-state electrochromic device. *Nano Lett.* **19**, 7988–7995 (2019).

39. Chen, S. & Jonsson, M. P. Dynamic conducting polymer plasmonics and metasurfaces. *ACS Photonics* **10**, 571–581 (2023).

40. Lu, W., Jiang, N. & Wang, J. Active electrochemical plasmonic switching on polyaniline-coated gold nanocrystals. *Adv. Mater.* **29**, 1604862 (2017).

41. Lu, W., Chow, T. H., Lai, S. N., Zheng, B. & Wang, J. Electrochemical switching of plasmonic colors based on polyaniline-coated plasmonic nanocrystals. *ACS Appl. Mater. Interfaces* **12**, 17733–17744 (2020).

42. Lu, Y. et al. All-state switching of the Mie resonance of conductive polyaniline nanospheres. *Nano Lett.* **22**, 1406–1414 (2022).

43. Decker, M. et al. High-efficiency dielectric Huygens' surfaces. *Adv. Opt. Mater.* **3**, 813–820 (2015).

44. Lu, W. Chow, T. H., Lu, Y. & Wang, J. Electrochemical coating of different conductive polymers on diverse plasmonic metal nanocrystals. *Nanoscale* **12**, 21617–21623 (2020).

45. Pensa, E. et al. Spectral screening of the energy of hot holes over a particle plasmon resonance. *Nano Lett.* **19**, 1867–1874 (2019).

46. Pud, A. A. Stability and degradation of conducting polymers in electrochemical systems. *Synth. Met.* **66**, 1–18 (1994).

47. Lacroix, J. C., Kanazawa, K. K. & Diaz, A. Polyaniline: a very fast electrochromic material. *J. Electrochem. Soc.* **136**, 1308–1313 (1989).

48. Heinze, J., Frontana-Uribe, B. A. & Ludwigs, S. Electrochemistry of conducting polymers–persistent models and new concepts. *Chem. Rev.* **110**, 4724–4771 (2010).

## Methods

### Numerical simulation

For the numerical simulation of the active Huygens' nanoantenna, we employed a finite-difference time-domain solver (Ansys Lumerical) using periodic boundary conditions in the *x* and *y* directions. The silicon material data took from Palik[49] and the polymer index took from the experimentally measured results obtained through ellipsometry (Supplementary Fig. 1). For the geometric parameters, the Si nanodisks have a height of 140 nm and a varying radius from 60 nm to 160 nm, and the height of polymer is 200 nm. Transmission spectra, phase and transmittance of the nanoantenna were simulated under a plane-wave excitation with a linear polarization and wavelength range from 500 nm to 1000 nm.

### Metasurface fabrication

The dielectric metasurfaces were fabricated by a nanofabrication procedure that comprises EBL patterning, mask deposition and RIE. In detail, a 140-nm thick Si film were deposited on a transparent conductive 50-nm thick ITO-coated fused silica glass substrate using a plasmon-enhanced chemical vapor deposition. For the lithography, a double layered poly(methyl-methacrylate) (PMMA) positive photoresist (495k A4 and 950k A2) was spin-coated onto the Si film with a soft-baking for 90 s at 170 °C, followed by spin-coating of a conducting layer (ESpace 300Z) on the photoresist to avoid electron charge and pattern distortion. The array of Si nanodisks was patterned on the photoresist using EBL (Raith eLine Plus), followed by a development process by immersing the sample into a 3:1 Isopropanol:Methylisobutylketone solution for 50 s. A 30-nm thick chromium layer was deposited using an e-beam evaporation as a hard mask. Lift-off process was carried out in a remover solution (Microposit Remover 1165). The designed pattern was finally etched into the Si film by a RIE process (Oxford Instrument) for 2 min. As the last step, the chromium hard mask was chemically removed by wet etching using a chromium-selective etchant solution (Sigma-Aldrich).

**Electrochemical setup**

The electrochemical polymer growth and electrical switching were carried out in a specially custom-built electrochemical cell. The electrochemical cell is designed for housing a three-electrode system into a thin layer of aqueous electrolyte with an optical thickness of 1 mm on the top of the sample substrate, where a thin transparent glass was used to seal the electrolyte on the top, allowing for an optical transmission measurement through the metasurface. The sample substrate (ITO-coated fused silica) was used as the bottom sealing glass of the cell. The cell also features on the side the in-let and out-let for electrolyte to enable flow-in and flow-out for electrolyte replacement, which is driven by an electrical injector. The sample substrate was connected as the working electrode through a striped metal plate as the electrode contact, whereas a Pt wire and a Ag/AgCl reference were inserted from the side of the cell and connected as the counter electrode and the reference electrode, respectively. A potentiostat (CHI-760e) is used to apply voltage over the time to perform electrochemical polymer growth and switching.

**In-situ polymer growth and electrical switching**

In-situ PANI growth was carried out by an electrochemical coating method, as reported previously[44]. The ITO substrate with the metasurface sample was connected as the working electrode. A cycling voltage in the range from -0.2 V to +0.8 V at a scanning speed of 25 mV/s was applied on the sample in an acidic aqueous electrolyte containing 1 M $H_2SO_4$ and 0.2 M aniline. The thickness of the grown PANI thickness can be controlled by the number of voltage cycle.

For the electrical switching, the electrolyte in the electrochemical cell was replaced by an aniline-free aqueous electrolyte containing only 1 M $H_2SO_4$. To switch the PANI to the oxidized state and reduced state, constant voltages of +0.6 V and -0.2 V, respectively, were applied on the substrate. For electrical switching cycling, a cycling voltage in the range between -0.2 V and +0.6 V at a scanning speed of 25 mV/s was used.

**Structural characterization**

Scanning electron microscopy (SEM) of the metasurface was performed on the Raith eLine Plus system in a SEM mode under a working voltage of 5 kV. The height of the deposited Si film and the thickness of the grown PANI layer at the optimized coating cycle were measured with a profilometer (Bruker Dektak XT) using a stylus with a radius of 2 μm.

**Optical characterization**

The refractive index of PANI was obtained from an ellipsometry measurement on an ellipsometer with dual-rotating compensators and a spectrometer (J.A. Woollam, M2000XI-210). A 100-nm thick PANI film electrochemically prepared on an ITO-coated glass substrate was measured with angle-variable spectroscopic ellipsometry at incident angles of 65°, 70° and 75°, using a bare ITO-coated glass as a blank reference. The measured refractive index was extracted from an experimentally fitted oscillator model based on a previous study[50].

Transmission spectra were taken using a commercial white light transmission microscopy setup (Witec Alpha series 300). The homogeneous metasurface sample was illuminated by a normal incident collimated white light with a linear polarization. The transmitted light was collected using a 20x objective with NA = 0.4 and direct to a grating-based spectrometer.

Diffraction patterns of the transmitted light was collected on a home-built optical setup as shown in Supplementary Figure 4. A white light source as well as a 785 nm laser were coupled into the optical path as light sources. The custom-built electrochemical cell with the metasurface sample was placed normally to the incident laser, where a 20x objective with NA = 0.4 was used to collect the optical response of the metasurface. A high-speed camera with a monochromatic charge-coupled device (Basler acA1440-220um) was employed to locate the sample and monitor the diffraction patterns.

**References**

49. Palik, E. D. *Handbook of Optical Constants of Solids*. (Academic Press, 1998).

50. Al-Attar, H. A., Al-Alawina, Q. H. & Monkman, A. P. Spectroscopic ellipsometry of electrochemically prepared thin film polyaniline. *Thin Solid Films* **429**, 286–294 (2003).

**Data Availability**

All the relevant data that support the findings of this study are available from the corresponding author upon reasonable request.

**Acknowledgements**

This work was funded by the Deutsche Forschungsgemeinschaft (DFG, German Research Foundation) under Germany's Excellence Strategy (EXC 2089/1 – 390776260), MA 4699.7-1, and the Emmy Noether program (TI 1063/1), the Bavarian program Solar Energies Go Hybrid (SolTech), and the Center for NanoScience (CeNS), Ludwig-Maximilians-University Munich. W.L. acknowledges the funding support of the Humboldt Research Fellowship from the Alexander von Humboldt Foundation. H.R. acknowledges the funding support from the Australian Research Council (DECRA Project DE220101085). H.R. and S.A.M. acknowledge the funding support from the Australian Research Council (Discovery Project DP220102152). S.A.M additionally acknowledges the Lee-Luces Chair in Physics.

**Author contributions**

W.L and S.A.M. conceived the project. W.L performed the simulation and designed the metasurface and carried out the nanofabrication. L.de.S.M. and W.L. constructed the optical measurement setup. W.L. designed the customized electrochemical cell and conducted the electrical switching experiments. A.T. and H.R. contributed to the metasurface design and made helpful comment on the manuscript. S.A.M. offered ideal facility support for carrying out the experiments. All authors discussed the results, analyzed the data and commented on the manuscript.

**Competing interests**

The authors declare no competing interests.

# Active Huygens' metasurface based on in-situ grown conductive polymer

# Supplementary Information

*Wenzheng Lu[1,*], Leonardo de S. Menezes[1,2,3], Andreas Tittl[1], Haoran Ren[4] and Stefan A. Maier[4,5,*]*

[1]Chair in Hybrid Nanosystems, Nano-Institute Munich, Faculty of Physics, Ludwig-Maximilians-Universität München, Munich, 80539, Germany.

[2]Departamento de Física, Universidade Federal de Pernambuco, 50670-901 Recife-PE, Brazil.

[3]Center for Nanoscience, Faculty of Physics, Ludwig-Maximilians-Universität München, Munich, 80539, Germany.

[4]School of Physics and Astronomy, Monash University, Clayton, Victoria 3800, Australia.

[5]Department of Physics, Imperial College London, London SW72AZ, UK.

[*]e-mail: Wenzheng.Lu@physik.uni-muenchen.de; Stefan.Maier@monash.edu

# Supplementary information for

# Active Huygens' metasurface based on in-situ grown conductive polymer

*Wenzheng Lu[1,]\*, Leonardo de S. Menezes[1,2,3], Andreas Tittl[1], Haoran Ren[4] and Stefan A. Maier[4,5,]\**

**Table of contents**

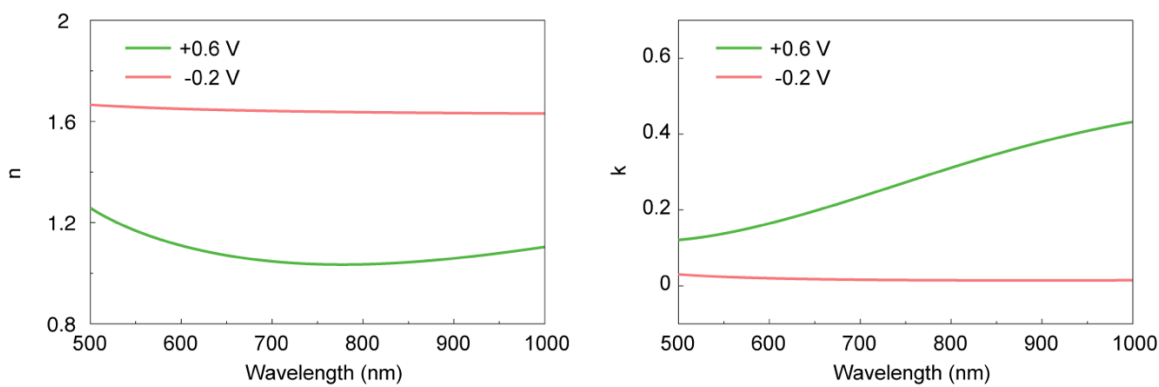

**Supplementary Fig. 1 | Refractive index of polyaniline.** Real part (left) and imaginary part (k) of refractive index of polyaniline (PANI) in the oxidized state (+0.6 V, green line) and the reduced state (-0.2 V, red line), respectively. The refractive index was experimentally measured by an ellipsometer on a 100-nm thick electrochemically prepared PANI film on an indium-tin-oxide (ITO)-coated fused silica substrate.

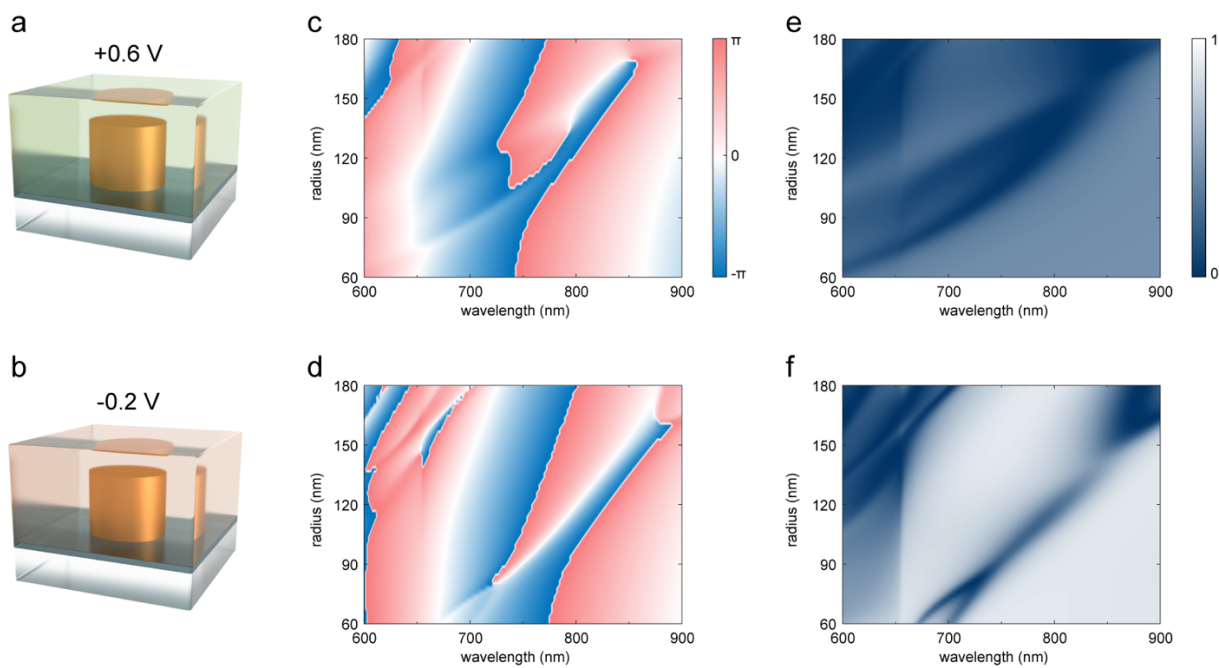

**Supplementary Fig. 2 | Simulation of electrically active Huygens' nanoantennas. a,b**, Schematics of an individual nanoantenna made of a silicon nanodisk surrounded by a layer of

PANI at an applied voltage of +0.6 V and -0.2 V, respectively. **c,d**, Simulated transmission optical phase profiles (color-coded) of the nanoantenna at +0.6 V and -0.2 V, respectively. **e,f**, Simulated transmittance (color-coded) of the nanoantenna at +0.6 V and -0.2 V, respectively.

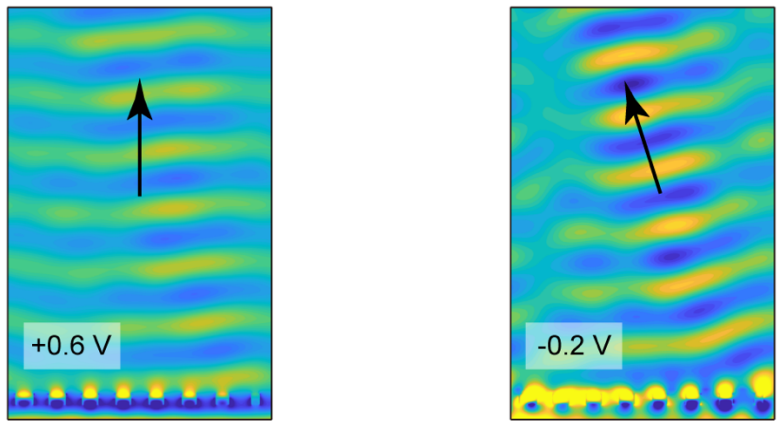

**Supplementary Fig. 3 | Simulated electric field distribution.** Simulated electric field distribution of the transmitted light for the electrically active metasurface under a normal incident at an applied voltage of +0.6 V (left) and -0.2 V (right), respectively. The arrow indicates the k vector of the transmitted light propagating wave.

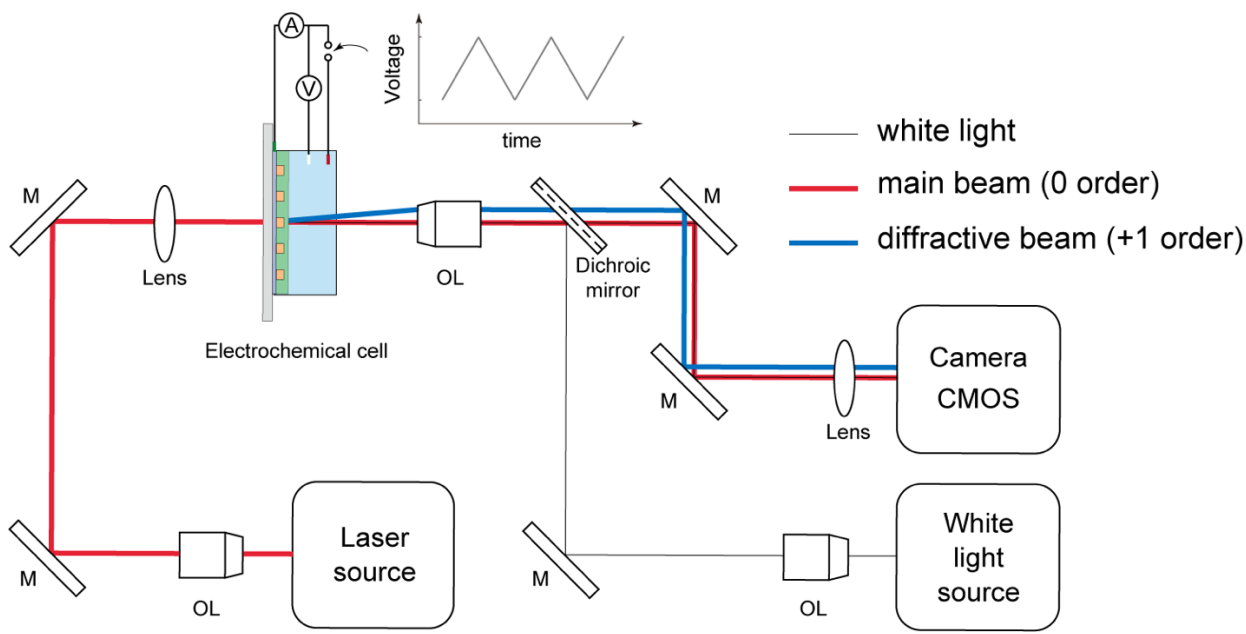

**Supplementary Fig. 4 | Optical measurement setup.** Schematic of the optical measurement setup integrated with the custom-built electrochemical cell for measuring the transmitted intensities of different diffraction orders. The white light source is used for locating the metasurface sample, and is turned off when performing optical measurement. During the polymer growth and the electrical switching, a 785-nm laser is used to illuminate at the center of the metasurface, while a cycling voltage is applied on the metasurface to induce variation of transmitted diffraction pattern. The transmitted diffraction pattern is recorded by a high-speed monochromatic camera CMOS. OL: objective lens, M: plane mirror.

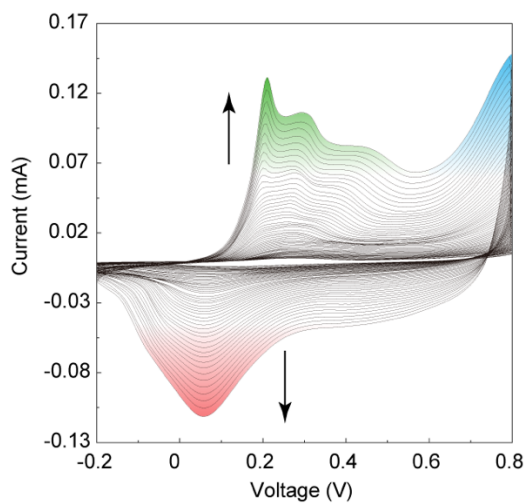

**Supplementary Fig. 5 | Cyclic voltammogram for electrochemical polymer growth.** The measured cyclic voltammogram for the electrochemical growth of PANI on the dielectric metasurface, including 60 coating cycles. The green area shows the region where the PANI oxidation process takes place, whereas the red area shows the reduction process. The blue area shows the oxidative polymerization process of aniline, resulting in an increasing PANI thickness (see also supplementary note 1). The arrow indicates the increasing current intensity as more PANI is grown on the metasurface.

**Supplementary Note 1 | In-situ electrochemical polymer growth on dielectric metasurfaces.**
In the presence of the precursor monomer, in this case, aniline, in the electrolyte, application of a high voltage can trigger an oxidative polymerization on the surface of the ITO substrate directly from the electrolyte[1]. The dielectric metasurfaces resting on the top of the ITO substrate exposed to the electrolyte can thus be grown with PANI. The voltage for oxidative polymerization of aniline in our case is approximately +0.8 V (vs. Ag/AgCl reference), as depicted by the rising current intensity in the blue area of the cyclic voltammogram in Supplementary Figure 5. On the other hand, PANI can be switched between the oxidized state and the reduced state by an applied voltage in the range from +0.6 V to -0.2 V. Therefore, a cycling voltage in the range between -0.2 V and +0.8 V is able to induce polymer growth while switching the PANI after each increment in the PANI thickness, thus allowing for an in-situ optimization of PANI thickness on the beam steering performance. The oxidation and reduction processes of PANI are recorded by the oxidation peak and reduction peak on the cyclic voltammogram. The increasing amount of PANI grown on the metasurface results in an increasing oxidization current and reduction current, as displayed on the green and red area of cyclic voltammogram in Supplementary Figure 5.

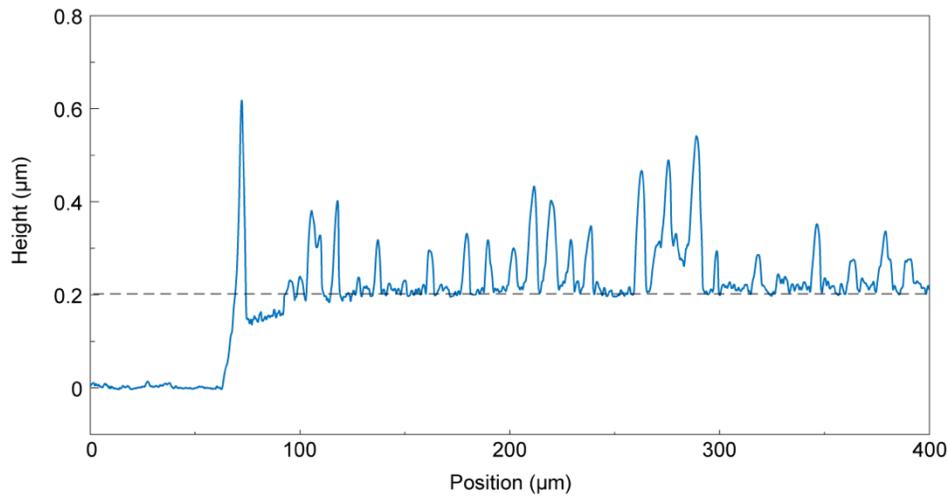

**Supplementary Fig. 6 | Measurement of PANI layer thickness.** Thickness profile of the PANI layer grown on the metasurface. The PANI layer is optimized at a coating cycle number of 56, which has a thickness of approximately 200 nm. The thickness of the PANI layer is measured by a profilometer. The appeared spikes in the profile is caused by the polymer surface roughness.

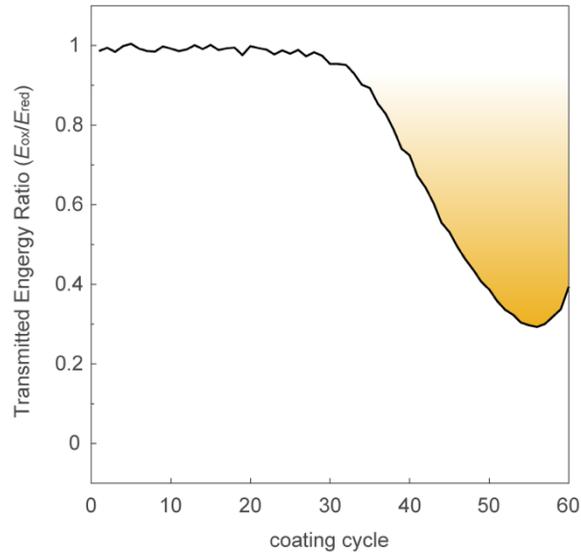

**Supplementary Fig. 7 | Transmitted energy contrast.** Transmitted energy ratio between the oxidized state and the reduced state ($E_{ox}/E_{red}$) at different polymer coating cycles. The transmitted energy is calculated from the sum of the transmitted intensities of +1, 0 and -1 diffraction orders.

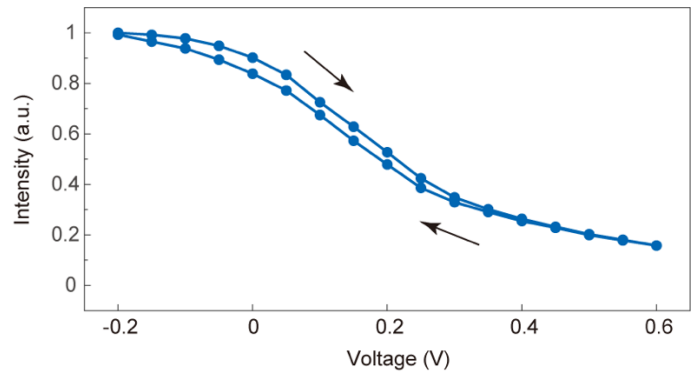

**Supplementary Fig. 8 | Hysteresis behavior of electrical switching.** Diffraction intensity of the +1 order at different applied voltage switched from the oxidized state and the reduced state. The arrows indicate the electrical switching direction. The narrow gap between the intensity of the two switching directions shows that the electrical switching of the PANI-integrated metasurface is nearly hysteresis-free, allowing for the accurate electrical controllability over the intermediate states.

**Supplementary Video 1 | Real-time electrical switching of the active Huygens' metasurface.** Images of the transmitted diffraction pattern during the electrical switching for 9 complete cycles. The video including the real-time applied voltage (top left), the cyclic voltammogram (top right), the intensity of +1 diffraction order (bottom left) and the intensity of 0 diffraction order (bottom right). The slightly spike-like signal of the 0 order diffraction intensity is caused by the dynamic variation of absorption coefficient of PANI when the applied voltage reverses at around +0.6 V, which is observed in other PANI-based demonstration[2]. The video runs at 30 fps.

**References**

1. Lu, W., Chow, T. H., Lu, Y. & Wang, J. Electrochemical coating of different conductive polymers on diverse plasmonic metal nanocrystals. *Nanoscale* **12**, 21617–21623 (2020).

2. Kaissner, R. et al. Electrochemically controlled metasurfaces with high-contrast switching at visible frequencies. *Sci. Adv.* **7**, eabd9450 (2021).